# Realization of epitaxial thin films of the superconductor K-doped BaFe$_2$As$_2$

Dongyi Qin,[1] Kazumasa Iida,[2] Takafumi Hatano,[2] Hikaru Saito,[3,4] Yiming Ma,[4] Chao Wang,[5] Satoshi Hata,[4,5] Michio Naito,[1] and Akiyasu Yamamoto[1]

[1]*Department of Applied Physics, Tokyo University of Agriculture and Technology, Koganei, Tokyo 184-8588, Japan*
[2]*Department of Materials Physics, Nagoya University, Chikusa-ku, Nagoya 464-8603, Japan*
[3]*Institute for Materials Chemistry and Engineering, Kyushu University, Kasuga, Fukuoka 816-8580, Japan*
[4]*Interdisciplinary Graduate School of Engineering Sciences, Kyushu University, Kasuga, Fukuoka 816-8580, Japan*
[5]*The Ultramicroscopy Research Center, Kyushu University, Nishi-ku, Fukuoka 819-0395, Japan*

The iron-based superconductor Ba$_{1-x}$K$_x$Fe$_2$As$_2$ is emerging as a key material for high magnetic field applications owing to the recent developments in superconducting wires and bulk permanent magnets. Epitaxial thin films play important roles in investigating and artificially tuning physical properties; nevertheless, the synthesis of Ba$_{1-x}$K$_x$Fe$_2$As$_2$ epitaxial thin films remained challenging because of the high volatility of K. Herein, we report the successful growth of epitaxial Ba$_{1-x}$K$_x$Fe$_2$As$_2$ thin films by molecular-beam epitaxy with employing a combination of fluoride substrates (CaF$_2$, SrF$_2$, and BaF$_2$) and a low growth temperature (350–420ºC). Our epitaxial thin film grown on CaF$_2$ showed sharp superconducting transition at an onset critical temperature ($T_c$) of 36 K, slightly lower than bulk crystals by ~2 K due presumably to the strain effect arising from the lattice and thermal expansion mismatch. Critical current density ($J_c$) determined by the magnetization hysteresis loop is as high as 2.2 MA/cm$^2$ at 4 K under self-field. In-field $J_c$ characteristics of the film are superior to the bulk crystals. The realization of epitaxial thin films opens opportunities for tuning superconducting properties by epitaxial strain and revealing intrinsic grain boundary transport of Ba$_{1-x}$K$_x$Fe$_2$As$_2$.

## I. INTRODUCTION

Iron-based superconductors (IBSCs) have been regarded as new candidate materials for superconducting applications [1]. There are several families including: $Ln$FeAsO ($Ln$-1111, $Ln$ = lanthanoid elements) with the highest superconducting transition temperature ($T_c$) of 55 K [2], $Ae$Fe$_2$As$_2$ ($Ae$-122, $Ae$ = alkaline-earth elements) with the highest $T_c$ of 38 K [3], and Fe$Ch$ (Fe$Ch$-11, $Ch$ = chalcogens) with $T_c$ ~ 15 K [4] but reaching above 40 K in thin films [5−9]. Among these materials, Ba-122 is one of the most promising materials for high magnetic field applications because of its high upper critical field ($H_{c2}$) with small anisotropy [10−12], large critical current density ($J_c$) [13−16], and large critical grain-boundary angle [17−19]. In fact, powder-in-tube processed K-doped $Ae$-122 wires and tapes [20,21], and K-doped Ba-122 bulk magnets [22] have been fabricated as proof-of-principle studies for high-field applications, although the information on the $J_c$ transparency across the grain boundary for K-doped $Ae$-122 is not clear yet. This is a sharp contrast to other IBSCs, where high-quality films have been grown on a variety of single- and bicrystalline, and technical substrates and, hence, fundamental and application-related research has been developed [23−25].

The difficulty of growing epitaxial $Ae_{1-x}$K$_x$Fe$_2$As$_2$ thin films is mainly due to the volatility of K. To overcome this problem, several attempts have been reported to date. Lee *et al.* reported a postannealing technique [26,27], where precursor thin films of BaFe$_2$As$_2$ were postannealed in a quartz tube with K metal. The resultant films are *c*-axis oriented and show a superconducting transition with $T_c^{on} - T_c^{end}$ = 40.0–37.5 K; however, they were not epitaxially grown. More recently, Hatakeyama *et al.* and Hiramatsu *et al.* have developed a solid-phase epitaxy technique to prepare epitaxial thin films of the end member, KFe$_2$As$_2$ [28,29]. To produce KFe$_2$As$_2$ films, it is essential to suppress revaporization of K at high annealing temperatures, *e.g.*, as high as 1000°C.

Previously we have reported the importance of the low-temperature growth for incorporating volatile K into Sr$_{1-x}$K$_x$Fe$_2$As$_2$ and Ba$_{1-x}$K$_x$Fe$_2$As$_2$ films [30−33]. Films grown on bare oxide substrates were polycrystalline and *c*-axis oriented films were realized only with (K,As)-buffered substrates. These *c*-axis oriented films were not in-plane aligned but the highest $T_c$ of 38.3 K was attained. Hence, a proper substrate should be explored for realizing the epitaxial growth of K-doped Ba-122. Fluoride substrates (CaF$_2$, SrF$_2$, and BaF$_2$) are typically superior to oxide substrates for thin-film growth of IBSCs [34−38]. Additionally, even for a wide range of the growth temperature (280−500°C), epitaxially grown Fe(Se,Te) thin films are realized [34,39]. Here we employ the combination of a low-temperature growth, up to 420°C, and fluoride substrates that yield truly epitaxial films of Ba$_{1-x}$K$_x$Fe$_2$As$_2$ with a high superconducting transition ($T_c^{on}$ ~ 36 K), a sharp transition width of 1.5 K, and a high self-field $J_c$ of 2.2 MA/cm$^2$ at 4 K.

## II. EXPERIMENT

The Ba$_{1-x}$K$_x$Fe$_2$As$_2$ thin films were grown in a custom-designed molecular-beam epitaxy (MBE) chamber (base pressure of ~1 × 10$^{-9}$ Torr). The details of our film growth approach have been presented in our previous report [23]. Briefly, all elements (Ba, Fe, and As) except for K were supplied from pure metal sources by resistive heating. Elemental K was supplied from an In–K alloy (In$_8$K$_5$). We used electron impact emission spectrometry (EIES) [23] and atomic absorption spectrometry (AAS) [23] for the real-time rate monitoring of various elements: EIES for Ba and Fe, and AAS for K. The K-doping level in Ba$_{1-x}$K$_x$Fe$_2$As$_2$ films was aimed at $x$ = 0.4. The amount of supplied K flux was substantially (2–3 times) higher than the nominally required amount to compensate for K reevaporation. The As flux was optimized by adjusting the cell temperature. The growth rate was ~1.5 Å/s and the deposition time of 10 min yielded Ba$_{1-x}$K$_x$Fe$_2$As$_2$ thin films ~1000Å thick. The optimum growth temperatures were

700–720°C for parent BaFe$_2$As$_2$ and 350–420°C for Ba$_{1-x}$K$_x$Fe$_2$As$_2$. Fluoride substrates, CaF$_2$(001), SrF$_2$(001), and BaF$_2$(001), were used in this study. MgO(001) substrate was used as a reference. K-containing films, especially grown on oxide substrates, quickly degraded in air; therefore, we coated films with a polystyrene resin (commercially called "Q-dope") diluted by toluene as a capping layer immediately after removing the films from the MBE chamber.

The films were characterized by *in situ* reflection high-energy electron diffraction (RHEED), x-ray diffraction (XRD) using both two-circle ($\theta, 2\theta$) and four-circle ($2\theta, \omega, \phi, \chi$) diffractometers with Cu-$K_\alpha$ radiation, transmission electron microscopy (TEM), and resistivity ($\rho$–$T$) measurements. Magnetic measurements were performed using a superconducting quantum interference device magnetometer to determine $T_c$ and $J_c$. $J_c$ values were determined by the Bean critical state model.

## III. RESULTS AND DISCUSSION

### A. Crystal structure

Figure 1 shows the XRD patterns and RHEED images of Ba$_{1-x}$K$_x$Fe$_2$As$_2$ films. In Fig. 1(a), the XRD patterns of Ba$_{1-x}$K$_x$Fe$_2$As$_2$ films grown on CaF$_2$ and MgO are compared. The film grown on MgO(001) does not show any appreciable diffraction peaks outside the (002) MgO peak, which indicates that the film is polycrystalline. On the other hand, the film grown on CaF$_2$ is essentially single phased and *c*-axis oriented with minor iron arsenide impurity peaks. Compared to Ba$_{1-x}$K$_x$Fe$_2$As$_2$ films, pristine BaFe$_2$As$_2$ films (single phased and *c*-axis oriented) were easily grown on either MgO or CaF$_2$. Figure 1(b) shows the result of x-ray in-plane measurements, *i.e.*, the $\phi$ scan of the (103) reflection peak of the Ba$_{0.6}$K$_{0.4}$Fe$_2$As$_2$ film on CaF$_2$. The fourfold symmetry observed in the $\phi$ scan with a small full width at half maximum value $\Delta\phi \sim 1.39°$ confirms that the film is in-plane aligned and single crystalline. Similar results were obtained in the films grown on SrF$_2$ and BaF$_2$, suggesting that the use of fluoride substrates is critical for obtaining single-crystalline Ba$_{1-x}$K$_x$Fe$_2$As$_2$ films (see Supplemental Material, Fig. S1 [40]). The lattice parameters of the film on CaF$_2$ are $a_f \sim$ 3.889Å and $c_f \sim$ 13.380Å. The literature [16,40] shows that the lattice parameters of bulk samples follow $a_b = 3.96 - 0.12x$ (Å) and $c_b = 13.00 + 0.84x$ (Å). Estimating on the basis of these bulk data, Ba$_{1-x}$K$_x$Fe$_2$As$_2$ films with $x = 0.40$ should have $a = 3.91$Å and $c = 13.33$Å. The actual film has shorter $a_f$ and longer $c_f$ compared to this estimation. The difference arises from the thermal expansion mismatch and/or the in-plane lattice mismatch between the Ca-Ca distance (3.863Å) of CaF$_2$ and the in-plane lattice constant (3.91Å) of Ba$_{0.6}$K$_{0.4}$Fe$_2$As$_2$. The in-plane compressive strain leads to the out-of-plane tensile strain, which leads to a stretch in $c_f$ arising from the Poisson effect.

As seen in Figs. 1(c) and 1(d), the RHEED patterns of pristine BaFe$_2$As$_2$ and Ba$_{1-x}$K$_x$Fe$_2$As$_2$ grown on CaF$_2$ demonstrate that both films are single crystalline, which agrees with the XRD results. There is a small difference between the two samples. Pristine BaFe$_2$As$_2$ shows streaks whereas Ba$_{1-x}$K$_x$Fe$_2$As$_2$ shows streaky spots, which indicates that transmission diffraction is partly involved. Specifically, the surface of Ba$_{1-x}$K$_x$Fe$_2$As$_2$ is less smooth than that of pristine BaFe$_2$As$_2$. However, immediately after starting the growth of Ba$_{1-x}$K$_x$Fe$_2$As$_2$, the diffraction pattern in the RHEED image starts with streaks. This observation suggests smooth initial growth of Ba$_{1-x}$K$_x$Fe$_2$As$_2$ on CaF$_2$, which indicates a good physical and chemical matching between the films and the substrate.

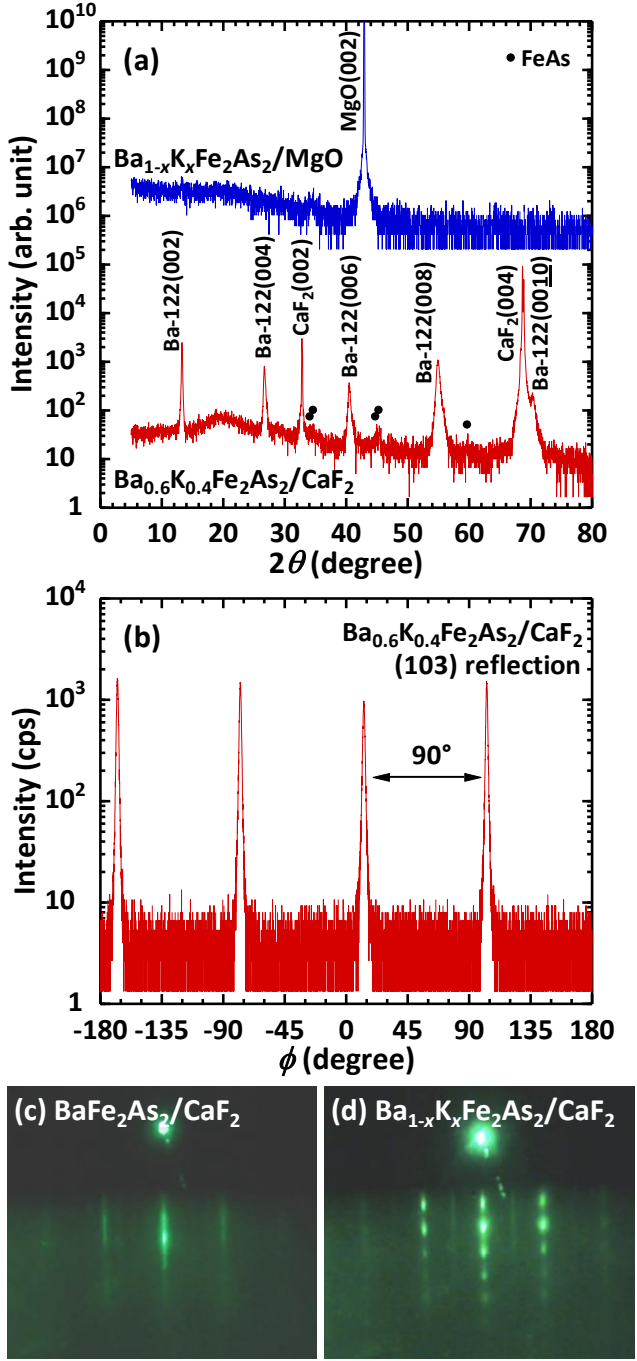

FIG. 1. (a) Comparison of the $\theta$–$2\theta$ XRD patterns of the Ba$_{1-x}$K$_x$Fe$_2$As$_2$ films on MgO (top) and CaF$_2$ substrates (bottom). (b) $\phi$ scan of the (103) reflection of the Ba$_{0.6}$K$_{0.4}$Fe$_2$As$_2$ film on CaF$_2$ substrate. (c), (d) RHEED patterns of the pristine BaFe$_2$As$_2$ (c) and Ba$_{1-x}$K$_x$Fe$_2$As$_2$ (d) films on CaF$_2$ substrates. The RHEED images were obtained after the growth with an incident electron beam parallel to the substrate [110] azimuth.

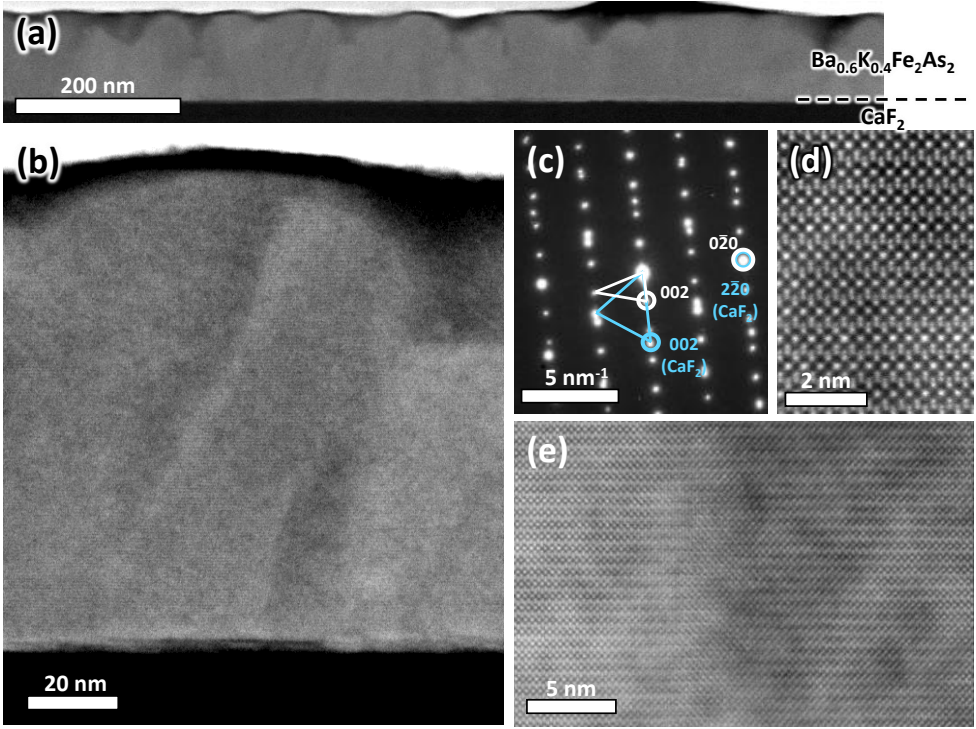

FIG. 2. Structural characterization by ADF-STEM imaging (at several different magnifications) and SEAD. (a), (b) Low- and medium-resolution ADF-STEM images. (c) Different pattern obtained from an area of a 100-nm size over the $Ba_{0.6}K_{0.4}Fe_2As_2$/$CaF_2$ interface. (d), (e) High-resolution ADF-STEM images obtained in a grain and at a grain boundary. Note that the top layer with the brightest contrast in (a) and (b) is a Pt layer deposited for protection against beam damage during the fabrication process with a focused ion beam.

### B. TEM analysis

Figure 2(a) shows the cross-sectional image of a $Ba_{0.6}K_{0.4}Fe_2As_2$ film grown on a $CaF_2$ substrate obtained by annular dark-field (ADF) scanning transmission electron microscopy (STEM). The cross-sectional sample was fabricated by a focused ion beam. The incident direction of the electron beam is parallel to the [110] direction of the $CaF_2$ substrate. The image contrast slightly varies in the $Ba_{0.6}K_{0.4}Fe_2As_2$ layer mainly in the in-plane direction, which indicates that the formation of columnar grains occurs perpendicular to the substrate surface. This is more clearly seen in a magnified image shown in Fig. 2(b), where several boundaries lie approximately perpendicular to the substrate surface at a several tens of nanometers interval. The columnar growth observed in TEM agrees with the RHEED image in Fig. 1(d). Of note, a lattice fringe is recognized in the entire area of the $Ba_{0.6}K_{0.4}Fe_2As_2$ layer in Fig. 2(b), which corresponds to the (001) interplanar spacing of $Ba_{0.6}K_{0.4}Fe_2As_2$. This means that the $Ba_{0.6}K_{0.4}Fe_2As_2$ film has strong (001) texture and almost no defects such as stacking fault or crack. Figure 2(c) shows a selected area electron diffraction (SAED) obtained from an area of 100-nm size over the $Ba_{0.6}K_{0.4}Fe_2As_2$/$CaF_2$ interface, which indicates a clear epitaxial relationship, i.e., the (001) planes of $Ba_{0.6}K_{0.4}Fe_2As_2$ are parallel to the (001) planes of $CaF_2$ and the [100] direction of $Ba_{0.6}K_{0.4}Fe_2As_2$ is parallel to the [110] direction of $CaF_2$. An atomic-resolution image is shown in Fig. 2(d), where each atomic column in the [100] direction can be recognized [42,43]. This column arrangement is blurred at grain boundaries owing to misorientation as shown in Fig. 2(e). However, a similar atomic-column arrangement is still recognized, meaning that the misorientation is not so large. This small misorientation is consistent with the XRD results shown in Fig. 1(b). Energy-dispersive x-ray spectroscopy (EDX) mapping at the ADF-STEM field of view revealed that the distribution of K in the $Ba_{1-x}K_xFe_2As_2$ film is nearly homogeneous with $x \sim 0.40 \pm 0.07$ (see Supplemental Material, Fig. S2 [40]).

### C. Superconducting properties

Figure 3(a) demonstrates the superconducting transition observed by in-plane resistivity measurements for a $Ba_{0.6}K_{0.4}Fe_2As_2$ film on $CaF_2$, which shows a sharp transition with $T_c^{on}$ of 36.4 K. Here $T_c^{on}$ is defined as the temperature at which the resistivity becomes 90% of the normal state. The $T_c^{on}$ values of $Ba_{1-x}K_xFe_2As_2$ films on fluoride substrates are slightly ($\sim$2 K) lower than bulk $T_c$ of 38 K [3]. The slightly lower $T_c$ can be considered as a result of the in-plane compressive strain. The in-plane lattice mismatch between $CaF_2$ and $Ba_{1-x}K_xFe_2As_2$ increases by decreasing temperature from the growth temperature (673 K) to the measurement temperatures (i.e., $4 \leq T \leq 40$ K). The in-plane strain is estimated to be around 2.4% at 40 K (see Supplemental Material, Fig. S3 [40]). It is reported that the in-plane compressive strain gives a negative impact on $T_c$ for $Ba_{1-x}K_xFe_2As_2$ ($x \geq 0.28$) [44], which also supports our results of slightly lower $T_c$ on $CaF_2$ substrate. The inset of Fig. 3(a) compares the temperature dependences of resistivity for the epitaxial film on $CaF_2$ and the polycrystalline film on MgO. The residual resistance ratio (RRR) is fairly high ($\sim$7) for the epitaxial film of $Ba_{0.6}K_{0.4}Fe_2As_2$ on $CaF_2$. This trend was also confirmed for the films on other fluoride substrates (see Supplemental Material, Fig. S1 [40]). However, the polycrystalline film on MgO shows a lower RRR value (2.7) with a broad superconducting transition, although a $T_c^{on}$ of is 34.5 K reasonably high. Figure 3(b) shows the temperature dependence of magnetic susceptibility ($\chi$–$T$), for the zero-field-cooled (ZFC) and field-cooled (FC) $Ba_{0.6}K_{0.4}Fe_2As_2$ film on $CaF_2$. The data were normalized to their absolute value at 4 K. The film shows a clear diamagnetic signal below 33.5 K, which approximately agrees with the temperature ($T_c^{zero}$), at which the resistivity becomes zero in transport measurements. Although the $\chi$–$T$ curve for ZFC of this film shows a broad transition, most of our films grown by the similar condition exhibited a $T_c$ of 33 K with a sharp transition (see Supplemental Material, Fig. S4 [40]). Hence, the lower $T_c$ of our films than that of single crystals is not associated with the extrinsic factors such as chemical homogeneity and crystalline imperfection but rather the strain effect discussed above.

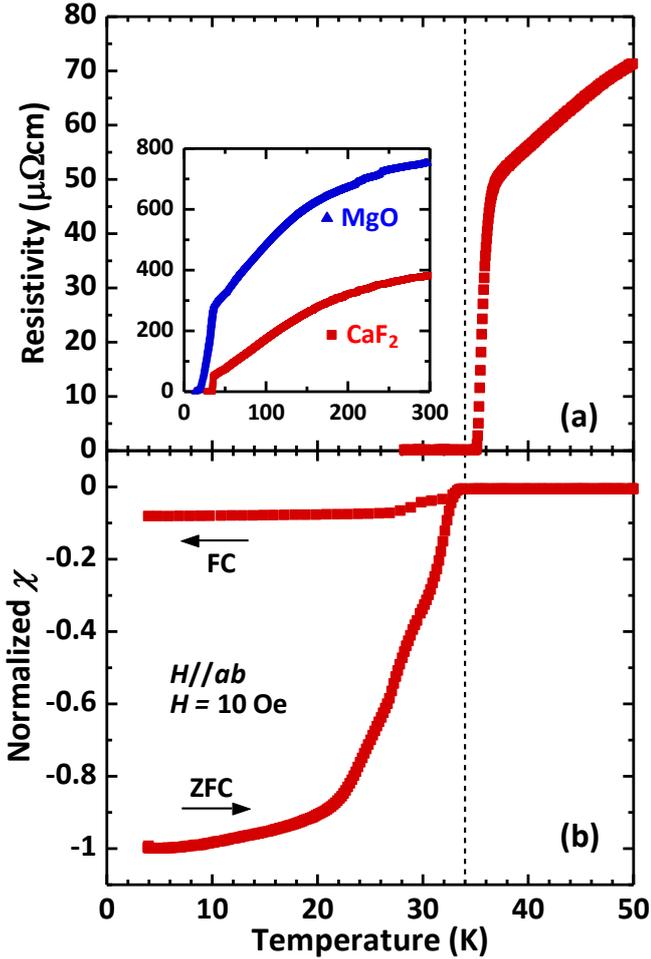

FIG. 3. Superconducting transitions measured by resistivity and susceptibility for $Ba_{0.6}K_{0.4}Fe_2As_2$ film on $CaF_2$ substrate. (a) Temperature dependence of resistivity near $T_c$. The inset shows comparison of the temperature dependences of resistivity for $Ba_{0.6}K_{0.4}Fe_2As_2$ films on $CaF_2$ and MgO substrates. (b) Temperature dependence of magnetic susceptibility of the $Ba_{0.6}K_{0.4}Fe_2As_2$ film on $CaF_2$ substrate for the ZFC and FC processes. The magnetic field was applied perpendicular to the crystallographic $c$-axis.

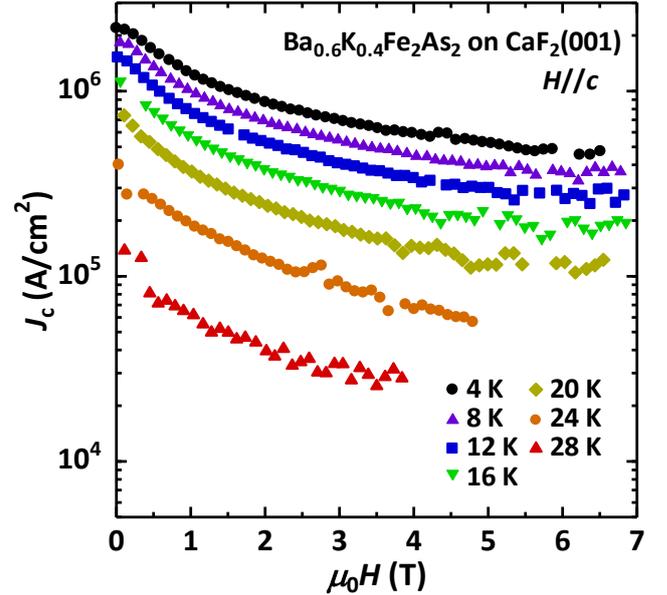

FIG. 4. Magnetic field dependence of $J_c$ for the $Ba_{0.6}K_{0.4}Fe_2As_2$ film on $CaF_2$ substrate. The magnetic field was applied parallel to the crystallographic $c$-axis.

Figure 4 shows the magnetic field dependence of $J_c$ for the $Ba_{0.6}K_{0.4}Fe_2As_2$ film on $CaF_2$. The self-field $J_c$ at 4 K is as high as 2.2 MA/cm$^2$ and decreases monotonously with the increase in temperature. However, the value is still above $10^5$ A/cm$^2$ even at 28 K. Of note, the level of $J_c$ for our thin films is superior to that of a single crystal [16]. The $J_c$ shows a rather weak dependence on the magnetic field, indicating that the $c$-axis correlated pinning centers are present in the film. As stated above, grain boundaries lie approximately perpendicular to the substrate surface at a several tens of nanometers interval, which may work as pinning centers. The $J_c$ values from the preliminary transport measurements were confirmed to be comparable to those from the magnetization measurements, indicating that supercurrent flows macroscopically in the film without being interrupted by dislocations or high-angle grain boundaries, which also supports that the sample is an epitaxial film. The detailed transport $J_c$ measurements of $Ba_{1-x}K_xFe_2As_2$ thin films would be the direction of our next study.

Our results expand the possibilities for wide-reaching researches. Firstly, the uniaxial pressure dependence of $Ba_{1-x}K_xFe_2As_2$ can be investigated, since in-plane strain can be induced by using the lattice or thermal expansion mismatch between films and fluoride substrates, e.g., Refs. [45,46]. Secondly and most importantly, single-crystalline-like $Ba_{1-x}K_xFe_2As_2$ thin films are realized for a wide range of substrates on which fluoride is epitaxially grown. This opens up avenues to grow $Ba_{1-x}K_xFe_2As_2$ on fluoride-buffered technical and bicrystalline substrates.

### IV. CONCLUSION

To conclude, we demonstrate the epitaxial growth of $Ba_{1-x}K_xFe_2As_2$ films on fluoride substrates such as $CaF_2$, $SrF_2$, and $BaF_2$. Both of the XRD and TEM analyses indicate that the films are highly textured and single-crystalline-like: the films consist of columnar nanograins showing small misorientation angles in the $a$-$b$ plane of $Ba_{1-x}K_xFe_2As_2$ at nanograin interfaces. The superconducting transition is as sharp as 1.5 K, although the $T_c^{on}$ of 36.4 K of $Ba_{0.6}K_{0.4}Fe_2As_2$ on $CaF_2$ is slightly (~2 K) lower than the bulk $T_c$. This $T_c$ suppression is due presumably to the strain effect. The self-field $J_c$ at 4 K is as high as 2.2 MA/cm$^2$ and decreases monotonously with an increase in temperature. It should be noted that the level of $J_c$ for our thin films is superior to that of a single crystal. $J_c$ shows a rather weak dependence on magnetic fields. The successful realization of epitaxial thin films opens the opportunities to investigate the transport properties for evaluating the application potentials.

## ACKNOWLEDGMENTS

The authors thank S. Takano (TUAT) for his early contribution to MBE growth of Ba$_{1-x}$K$_x$Fe$_2$As$_2$ on fluoride substrates. This work was partly supported by JST CREST (Grant No. JPMJCR18J4) and Advanced Characterization Platform of the Nanotechnology Platform Japan (Grant No. JPMXP09-A-19-KU-1003 and No. 1004) sponsored by the Ministry of Education, Culture, Sports, Science and Technology (MEXT), Japan.

# Realization of epitaxial thin films of the superconductor K-doped BaFe$_2$As$_2$: Supplemental Material

Dongyi Qin,[1] Kazumasa Iida,[2] Takafumi Hatano,[2] Hikaru Saito,[3,4] Yiming Ma,[4] Chao Wang,[5] Satoshi Hata,[4,5] Michio Naito,[1] and Akiyasu Yamamoto[1]

[1]*Department of Applied Physics, Tokyo University of Agriculture and Technology, Koganei, Tokyo 184-8588, Japan*
[2]*Department of Materials Physics, Nagoya University, Chikusa-ku, Nagoya 464-8603, Japan*
[3]*Institute for Materials Chemistry and Engineering, Kyushu University, Kasuga, Fukuoka 816-8580, Japan*
[4]*Interdisciplinary Graduate School of Engineering Sciences, Kyushu University, Kasuga, Fukuoka 816-8580, Japan*
[5]*The Ultramicroscopy Research Center, Kyushu University, Nishi-ku, Fukuoka 819-0395, Japan*

In this Supplemental Material, we include x-ray diffraction and temperature dependence of resistance data for Ba$_{1-x}$K$_x$Fe$_2$As$_2$ films grown on various fluoride (CaF$_2$, SrF$_2$, and BaF$_2$) substrates, chemical composition analysis by TEM-EDX, temperature dependence of the in-plane lattice parameter and in-plane strain, and superconducting transition temperature of several Ba$_{1-x}$K$_x$Fe$_2$As$_2$ thin films on CaF$_2$ substrates.

## I. X-ray diffraction and resistivity data for Ba$_{1-x}$K$_x$Fe$_2$As$_2$ films grown on various fluoride (CaF$_2$, SrF$_2$, and BaF$_2$) substrates

All the Ba$_{1-x}$K$_x$Fe$_2$As$_2$ films were grown simultaneously on CaF$_2$, SrF$_2$, and BaF$_2$ substrates with a slightly higher content of K compared to that of the film on CaF$_2$ substrate in Fig. 3 of the manuscript. Films are single crystalline, which is shown in Figs. S1 (a) and (b). All films are *c*-axis oriented in the $\theta$–$2\theta$ scan (a) and in-plane aligned in the $\phi$ scan (b). Temperature dependence of resistance measurements were also conducted for these three films and the data for the films on CaF$_2$ and SrF$_2$ are presented in Fig. S1 (c). Unfortunately, however, the film on BaF$_2$ was deteriorated before the *R*–*T* measurements, so the data is not presented.

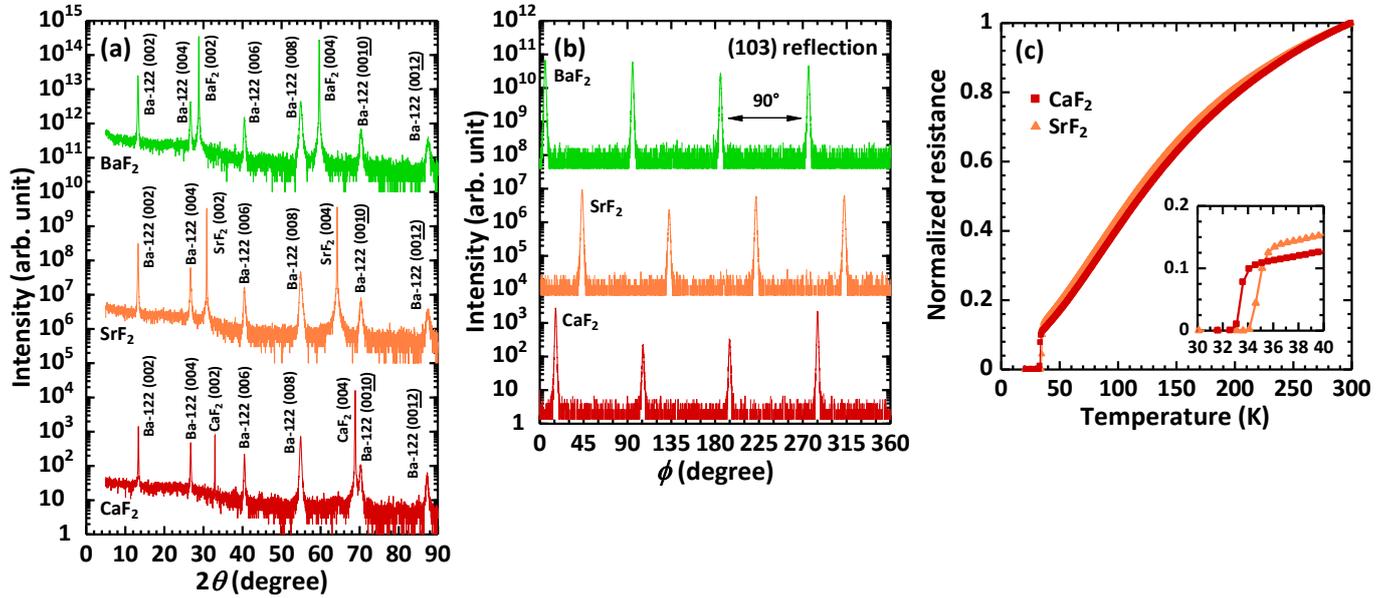

FIG. S1. (a) Comparison of the $\theta$–$2\theta$ XRD patterns of the Ba$_{1-x}$K$_x$Fe$_2$As$_2$ films on CaF$_2$, SrF$_2$, and BaF$_2$ substrates. (b) $\phi$ scan of the (103) reflection of the Ba$_{1-x}$K$_x$Fe$_2$As$_2$ films on CaF$_2$, SrF$_2$, and BaF$_2$ substrates. (c) Temperature dependences of normalized resistance for the Ba$_{1-x}$K$_x$Fe$_2$As$_2$ films on CaF$_2$ and SrF$_2$ substrates. The inset shows a comparison of the temperature dependences of normalized resistance near $T_c$.

## II. Chemical composition analysis by TEM-EDX

Chemical compositions in the grown film on the $CaF_2$ substrate were evaluated by energy-dispersive x-ray spectroscopy (EDX) performed in a transmission electron microscope (TEM). Elemental maps (characteristic x-ray intensity distributions) of Ba, K, Fe, As, and other related elements like Ca and F (substrate) are shown in Fig. S2 (a). The data were obtained from a cross-sectional slab sample extracted by a focused ion beam (FIB). After the FIB process, the cross-sectional slab sample was exposed in the air during sample transportation to the TEM, resulting in the detection of O atoms due to inevitable oxidation. The sample has a uniform thickness as understood from the flat contrast in the annular dark-field (ADF) image (top left). K atoms seem to distribute almost uniformly in the film, as shown in the elemental map, except for the vicinity of the substrate. The obtained EDX spectra were converted to chemical compositions by a conventional method using Cliff-Lorimer factors. Chemical composition variations along the normal to the substrate surface are plotted in Figs. S2 (b) and (c), where the compositions are averaged in the in-plane direction. In Fig. S2 (b), all the 8 elements are included in the composition calculation. Note that the x-ray absorption in the sample is not considered here, so the light elements, O and F, are severely underestimated. As shown, Ba, K, Fe, and As profiles are recognized to show a constant composition in a large part of the film, suggesting the formation of the stoichiometric film. A thin (~5 nm) layer of K-depleted $BaFe_2As_2$ is found at the interface between the film and the substrate, as denoted with a broken line and an arrowhead in Fig. S2 (c). This fact suggests that the substrate affects the film composition in the early stage of the film growth although the substrate effect can be negligible after 5 nm growth of the film.

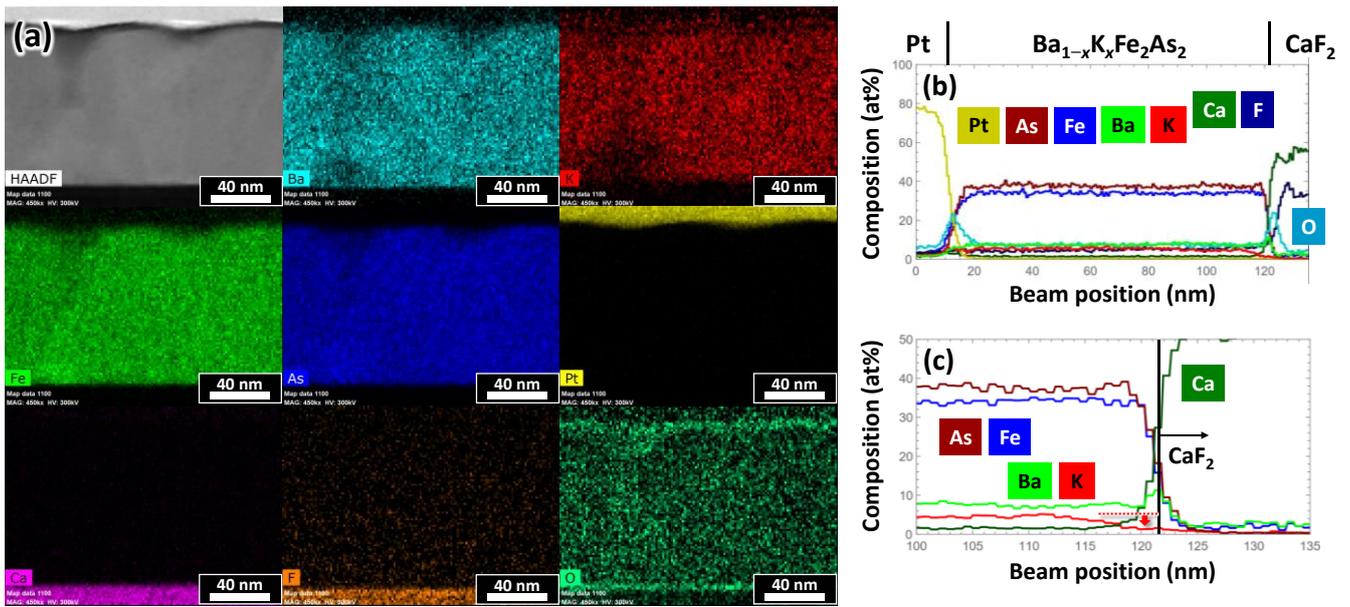

FIG. S2. (a) ADF-STEM image (top left) and corresponding elemental maps of Ba, K, Fe, As, Pt, Ca, F, and O taken by STEM-EDX. (b) Chemical composition variations along the normal to the substrate surface. The compositions of the plotted 8 elements were calculated by a conventional method using Cliff-Lorimer factors. (c) Magnified plots of (b) near the interface. Only the plots of 5 elements Ba, K, Fe, As, and Ca are displayed to avoid complexity.

### III. Temperature dependence of the in-plane lattice parameter and in-plane strain

The lattice parameters of both $CaF_2$ and $BaFe_{1.84}Co_{0.16}As_2$ as a function of temperature is shown in Fig. S3. The lattice parameters $a(T)$ were calculated by,

$$a(T) = a(300\text{ K}) \times \left[\exp\left\{\int_{300\text{ K}}^{T} \alpha(T) dT\right\}\right] \quad \left(\int_{300\text{ K}}^{T} \alpha(T) dT = \int_{a(300\text{ K})}^{a(T)} \frac{dl}{l}\right).$$

Using $a_{CaF_2}(300\text{ K}) = 5.463$ Å and the thermal expansion coefficient [1], $a_{CaF_2}(40\text{ K})$ is calculated to be 5.445 Å. The $\alpha(T)$ above 300 K for $Ba_{1-x}K_xFe_2As_2$ is not available, therefore we estimated the in-plane strain by using that of $BaFe_{1.84}Co_{0.16}As_2$ [2]. The in-plane strain at 40 K is calculated to be $\epsilon_{[100]} = \frac{a_{122}(40\text{ K}) - a_{sub.}(40\text{ K})}{a_{122}(40\text{ K})} \sim 0.024$. Note that $a_{sub.}(40\text{ K}) = \frac{a_{CaF_2}(40\text{ K})}{\sqrt{2}}$ since the basal plane of $Ba_{1-x}K_xFe_2As_2$ is rotated by 45º.

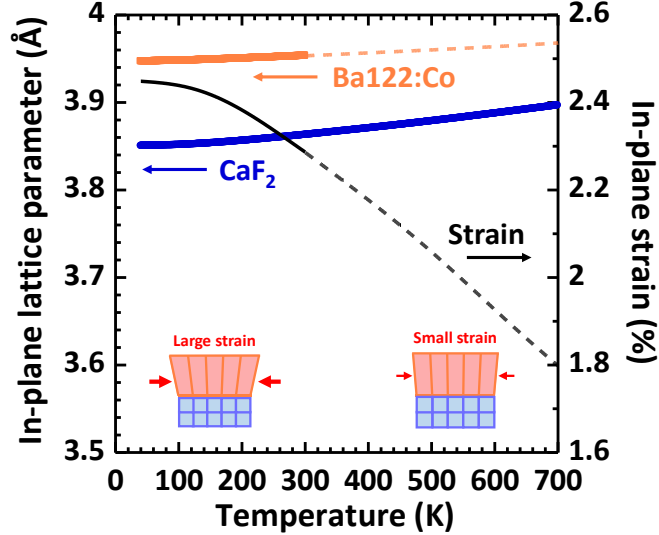

FIG. S3. Temperature dependence of the in-plane lattice parameters of $CaF_2$ and $BaFe_{1.84}Co_{0.16}As_2$ (left) and in-plane strain (right) calculated from refs. [1,2]. The dashed curves are a guide to the eye.

### IV. Superconducting transition temperature of $Ba_{1-x}K_xFe_2As_2$ thin films on $CaF_2$ substrates

Figure S4 shows the normalized $\chi$ for $Ba_{1-x}K_xFe_2As_2$ thin films on $CaF_2$ substrates as a function of temperature. The data is normalized to the value at 4 K. The growth condition was similar to the one presented in the manuscript. As can be seen, both films show a $T_c$ of 33 K with a sharp transition. Hence, the lower $T_c$ of our films than that of single crystals is not associated with the extrinsic factors such as chemical homogeneity and crystalline imperfection but rather the strain effect.

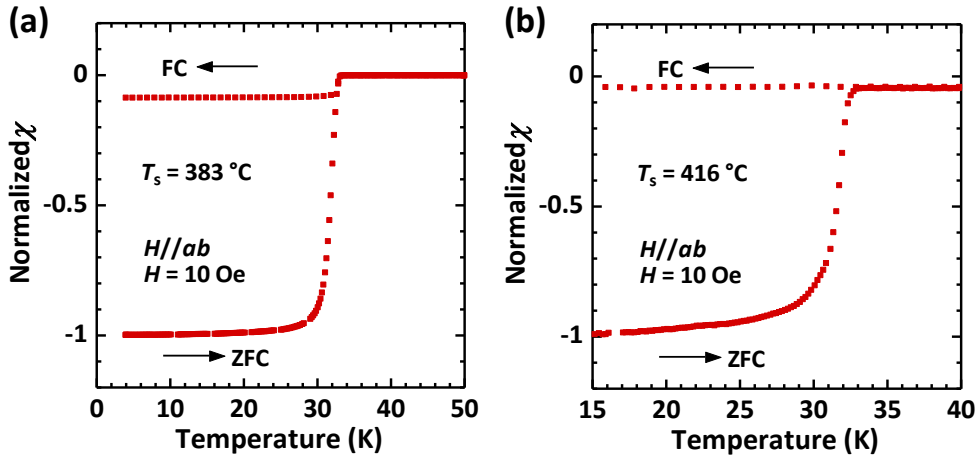

FIG. S4. Temperature dependence of the magnetic susceptibility for $Ba_{1-x}K_xFe_2As_2$ thin films grown on $CaF_2$. The growth condition is almost the same as presented in the manuscript. The growth temperature $T_s$ was (a) 383ºC and (b) 416ºC, respectively.